\documentclass[preprint,showpacs,showkeys]{revtex4}
\usepackage{amsmath,amsfonts,amssymb}

\begin{document}
\title{Equivalence principle and electromagnetic field: 
no birefringence, no dilaton, and no axion}
\author{Friedrich W. Hehl}
\email{hehl@thp.uni-koeln.de}
\affiliation{Institute for Theoretical Physics, University of Cologne, 
50923 K\"oln, Germany}
\altaffiliation[Also at: ]{Department of Physics \& Astronomy, 
University of Missouri-Columbia, Columbia, MO 65211, USA}

\author{Yuri N. Obukhov}
\email{yo@thp.uni-koeln.de}
\affiliation{Institute for Theoretical Physics, University of Cologne, 
50923 K\"oln, Germany\,}
\altaffiliation[Also at: ]{Department of Theoretical Physics, Moscow 
State University, 117234 Moscow, Russia}
\begin{abstract}
The coupling of the electromagnetic field to gravity is discussed. 
In the premetric axiomatic approach based on the experimentally well
established conservation laws of electric charge and magnetic flux,
the Maxwell equations are the same irrespective of the presence or
absence of gravity. In this sense, one can say that the charge 
``substratum" and the flux ``substratum" are not influenced by the 
gravitational field directly. However, the interrelation between
these fundamental substrata, formalized as the {\it spacetime
relation} $H=H(F)$ between the 2-forms of the electromagnetic excitation
$H$ and the electromagnetic field strength $F$, is affected by gravity.
Thus the validity of the equivalence principle for electromagnetism
depends on the form of the spacetime relation. We discuss the 
nonlocal and local linear constitutive relations and demonstrate that
the spacetime metric can be accompanied also by skewon, dilaton, and axion 
fields. All these premetric companions of the metric may eventually lead 
to a violation of the equivalence principle. 
\end{abstract}


\date{16 May 2007, {\it file equiv07.tex}}
\maketitle

\section{Introduction}

In the modern theory of gravity, one usually distinguishes three forms
of the equivalence principle. Namely, the {\it weak equivalence}
principle, relying on the experimental fact of the equality of the
gravitational and inertial masses, states that all massive objects
(test bodies) with the same initial velocity follow the same
trajectories in a gravitational field (the universality of a free
fall). The {\it Einsteinian equivalence} principle postulates that the
result of any local non-gravitational experiment in a local inertial
frame (``Einstein's elevator") is the same everywhere in spacetime.  A
generalization of this statement, when the same is true for any local
experiment, including the gravitational ones, is called the {\it
  strong equivalence} principle.

In gravitational theory, the question of whether the equivalence principle 
is valid and which form of it (as well as the interrelations between these 
forms), belongs to the most fundamental issues. Theoretical
analyses as well as experimental tests are conducted by many research
groups. Generally speaking, the equivalence principle specifies the
coupling of gravity to matter. Here we will not discuss all the aspects
of this subject, but we will confine our attention to the coupling between
gravity and electromagnetism. 

This paper is dedicated to our friend and colleague {\sl Bahram
  Mashhoon} on the occasion of his 60th birthday. One of us had quite
often discussions with Bahram on physics topics for over more than 30
years, and he is very grateful to Bahram for sharing with him his
profound insight into physics and astrophysics.

\section{Electromagnetism and equivalence principle}

Our world is made, to a great extent, of electromagnetically 
interacting particles. Moreover, practically all the information about
the near and the distant space comes to us in the form of electromagnetic
waves. Thus it is very important to verify whether the equivalence principle
is valid for electromagnetically interacting systems of fields and
particles. In the simplest formulation, we can ask: Does a photon move, 
since it is massless, along a null geodesic line  in a curved spacetime, 
in accordance with the equivalence principle? 

Within the framework of general relativity, a light ray can be
extracted from classical electrodynamics in its {\em geometrical
  optics} limit, i.e., for wavelengths much smaller than the local
curvature radius of space. Accordingly, the bending of light can be
understood as a result of a nontrivial refractive index of spacetime 
due to the coupling of the electromagnetic field $F$ to the gravitational 
field $g$, see Skrotskii et al.\ \cite{Skro1,Skro2}. Classically,
we have in nature just these two fundamental fields $F$ and $g$, the
weak and the strong fields being confined to microphysical dimensions
of $10^{-20}\, m$ or $10^{-15}\, m$, respectively. Therefore, the
coupling of $F$ and $g$ is of foremost importance in classical
physics.

\section{Premetric formulation of electrodynamics}

The equivalence principle is about the coupling of matter to gravity.
In order to understand the coupling of electromagnetism to gravity, it 
is most convenient and reasonable to start from the premetric formulation
of the electrodynamical theory. Following the early work of Kottler,
Cartan, and van Dantzig, one can indeed develop an axiomatic approach
to the electromagnetic field \cite{Birk,Zlatibor,Potsdam06} without 
assuming any specific geometric structure beyond the differentiable 
structure of the spacetime manifold, see also \cite{Punt,Gyros}. At the 
heart of this approach, there are the well established experimental facts 
(that can be formulated as fundamental axioms) of electric charge and 
magnetic flux conservation. The corresponding definitions and equations 
can be written down very succinctly in the calculus of {\em exterior 
differential forms} (Cartan calculus). Charge and flux conservation gives 
rise to the two basic objects of the electromagnetic theory, the 
electromagnetic excitation 2-form $H$ (with twist, or odd) and the 
electromagnetic field strength 2-form $F$ (without twist, or even), 
and it ultimately yields, with the electric current $J$, the field equations 
\begin{equation}\label{Maxwell}
dH=J\,,\qquad\qquad dF=0\,, \end{equation}
with 
\begin{equation}\label{ccexcalc}
dJ=0\,. \end{equation} 

The set (\ref{Maxwell}) represents the Maxwell equations. They are {\it
independent of metric and connection}. At this stage, the equivalence 
principle looks empty: Since the Maxwell equations (\ref{Maxwell}) are 
formulated in a coordinate and frame independent way, they are valid in 
this form in arbitrary coordinate systems and frames, be it in a flat or 
in a curved spacetime.

\section{Spacetime (vacuum constitutive) relation and equivalence principle}

In order to make electrodynamics a predictive theory, we need to 
complete Maxwell's equations (\ref{Maxwell}) with the {\it spacetime 
relation} (or {\it vacuum constitutive relation}). The latter links the 
excitation to the field strength,
\begin{equation}
H = H(F).\label{STrel}
\end{equation}
Only the constitutive relation ``feels'', up to a conformal factor, the 
presence of a flat or a nonflat metric $g$, i.e., the constitutive relation 
couples to the conformally invariant part of the metric. In addition,  
premetric companions can come into play through the spacetime relation 
(\ref{STrel}), as we will see later. The coupling of electromagnetism to 
gravity becomes almost trivial in this sense. In plain words, the test 
of the equivalence principle for electromagnetically interacting systems 
amounts to the test of the spacetime relation (\ref{STrel}).

The spacetime relation can be quite non-trivial, in general. 
For example, it can be nonlinear and even nonlocal. The nonlinearity can be 
fundamental or effective, see the discussion in \cite{Birk}. 
Here we will limit our attention to the linear spacetime relations. 

\subsection{Nonlocal, linear: Volterra-Mashhoon} 

Let us choose arbitrary local spacetime 
coordinates $x^i$ and the local coframe field $\vartheta^\alpha$. The latter
is usually determined by specifying the physical observers and their local
reference frames that they use for the making of physical (in particular
electrodynamical) measurements. Then we have with respect to the coframe 
$\vartheta^\alpha = e_i{}^\alpha\,dx^i$ for an in general non-inertial observer
\begin{equation}
H = {\frac 1 2}\,H_{\alpha\beta}\,\vartheta^\alpha\wedge\vartheta^\beta,
\qquad F = {\frac 1 2}\,F_{\alpha\beta}\,\vartheta^\alpha\wedge 
\vartheta^\beta.\label{non-localHF}
\end{equation}

Generalizing the locality assumption, Mashhoon \cite{Bahram} postulated, 
following similar suggestions of Volterra, the nonlocal constitutive law
\begin{equation}\label{non-local1}
H_{\alpha\beta}(\tau,\xi) = {\frac 1 2}\int d\tau' 
K_{\alpha\beta}{}^{\gamma\delta}(\tau,\tau')\,
F_{\gamma\delta}(\tau',\xi)\,.
\end{equation}
The response kernel in (\ref{non-local1}) is defined by the
acceleration and rotation of the observer's reference system. 
It is a constitutive law for the {\it vacuum} as viewed from a 
non-inertial frame of reference. 
The general form of the kernel can be worked out explicitly,
see \cite{Birk}, e.g., ($u$ is the observer's 4-velocity):
\begin{equation}
K_{\alpha\beta}{}^{\gamma\delta}(\tau,\tau') = {\frac 1
2}\,\epsilon_{\alpha\beta}{}^{\lambda[\delta}\!
\left(\delta^{\gamma]}_\lambda\,\delta(\tau -\tau') -
u\rfloor\Gamma_\lambda{}^{\gamma]}(\tau')\right).\label{NewAnsatz}
\end{equation}
The influence of non-inertiality is manifest in the presence of 
the connection 1-form $\Gamma_\alpha{}^\beta$. 
It has been shown that this kernel is the only consistent one, see the 
review \cite{Mashrev}. So far, there is no experimental support for
(\ref{non-local1}) and (\ref{NewAnsatz}) in the case of vacuum
electrodynamics. 

\subsection{Local, linear}

A very important case is that of a local and homogeneous linear constitutive 
law between the components of the two-forms $H$ and $F$. In local coordinates, 
we have $H = {\frac 1 2}\,H_{ij}\,dx^i\wedge dx^j, F = {\frac 1 2}\,F_{ij}
\,dx^i\wedge dx^j$. The linear spacetime relation then postulates the 
existence of a constitutive tensor with $6\times 6=36$ components 
$\kappa_{ij}{}^{kl}(t,x)= -\, \kappa_{ji}{}^{kl}=-\, \kappa_{ij}{}^{lk}$ 
such that
\begin{equation}
H_{ij} =  \frac{1}{2}\,\kappa_{ij}{}^{kl}\,F_{kl}.\label{chiHF}
\end{equation}
This kind of an ansatz we know from the physics of anisotropic crystals. 
Taking the Levi-Civita symbol, we can introduce the alternative constitutive 
tensor density by
\begin{equation}\label{lin1}
{\chi}^{ijkl} :=\frac{1}{2}\,\epsilon^{ijmn}\kappa_{mn}{}^{kl},\qquad 
{\rm or}\qquad \kappa_{ij}{}^{kl}=\frac{1}{2}\,\epsilon_{ijmn}\,{\chi}^{mnkl}.
\end{equation}
With the linear constitutive law (as with more general laws), we can,
in the case of vanishing dissipative effects, set up a Lagrangian 4-form; 
here we call it $V_{\rm lin}$.  Because of $H=-\,\partial V_{\rm lin}
/\partial F$, the Lagrangian must then be quadratic in $F$.  Thus we find
\begin{equation}\label{Vlin}
  V_{\rm lin} = -\,\frac{1}{2}\,H\wedge
  F=-\,\frac{1}{8}\,\chi^{ijkl}F_{ij}F_{kl}\,dx^0\wedge dx^1\wedge
  dx^2\wedge dx^3\,.
\end{equation}
The components of the field strength $F$ enter in a {\em symmetric} way.  
Therefore, within the framework of the variational approach, we can impose 
the symmetry condition $\chi^{ijkl} = \chi^{klij}$ on the constitutive 
tensor thus reducing the number of its independent components to 21 at this 
stage. The components $\kappa_{ij}{}^{kl}$ carry the dimension of $[\kappa]
=[{\chi}] = q^2/\mathfrak{h} =1/$resistance. We denote the dimension of
charge by $q$ and the dimension of action by $\mathfrak{h}$.

In this way, we come naturally to the so-called $\chi$-$g$ {\it scheme} 
proposed by Ni \cite{Nipreprint,Ni77,Ni84,Ni05} as a general phenomenological 
framework for the discussion of the equivalence principle. It includes  
narrower schemes such as \cite{Lightman73,Blanchet92,Denisov99}, e.g..

\section{Wave propagation}

How does an electromagnetic wave (a ``photon") propagate in space and time? 

\subsection{Generalized Fresnel equation} 

In the geometric optics approximation (equivalently, in the Hadamard 
approach) an electromagnetic wave is described by the propagation of 
a discontinuity of the electromagnetic field \cite{Yu+GuNonlinear}. 
The surface of discontinuity 
$S$ is defined locally by a function $\Phi$ such that $\Phi= const$ on $S$. 
The wave covector $q:=d\Phi$ contains the essential information about the 
propagation of electromagnetic waves (``light"). We define the 4th-order 
{T}amm--{R}ubilar (TR) tensor density of weight $+1$,
\begin{equation}\label{G4}
{\cal G}^{ijkl}(\chi):=\frac{1}{4!}\,{\epsilon}_{mnpq}\,
    {\epsilon}_{rstu}\, {\chi}^{mnr(i}\, {\chi}^{j|ps|k}\,
    {\chi}^{l)qtu }\,.
\end{equation} 
It is totally symmetric ${\cal G}^{ijkl}(\chi)= {\cal G}^{(ijkl)}(\chi)$.
Thus, it has 35 independent components. One can demonstrate that the wave
covectors satisfy the extended Fresnel equation that is generally covariant 
in 4 dimensions:
\begin{equation} \label{Fresnel} 
{\cal G}^{ijkl}(\chi)\,q_i q_j q_k q_l = 0 \,.  
\end{equation} 
In other words, the test of the equivalence principle for electromagnetic
waves reduces to the investigation of the structure of the wave surface 
(\ref{Fresnel}). The latter is determined solely by the spacetime relation,
i.e., by the constitutive tensor (\ref{lin1}) that enters the TR tensor
(\ref{G4}). New derivations of the extended Fresnel equation have been 
given recently by Itin \cite{ItinW} and Perlick \cite{PerlickW}.

\subsection{Birefringence, skewon} 

In general, the wave covectors $q$ lie on a 
{\em quartic Fresnel wave surface}, not exactly what we are observing in 
vacuum at the present epoch of our universe. One can gain further insight
into the wave propagation by decomposing the constitutive tensor into the
three irreducible parts \cite{PLA05,FP05}
\begin{eqnarray}\nonumber
  \kappa_{ij}{}^{kl} &=& {}^{(1)}\kappa_{ij}{}^{kl} +
  {}^{(2)}\kappa_{ij}{}^{kl} + 
  {}^{(3)}\kappa_{ij}{}^{kl} \\  &=&
{}^{(1)}\kappa_{ij}{}^{kl} - 4\!\not\!S_{[i}{}^{[k}\,\delta_{j]}^{l]} 
+ 2\,\alpha\,\delta_{[i}^k\delta_{j]}^l.\label{kap-dec}
\end{eqnarray}
The {\it skewon\/} and the {\it axion\/} fields are here conventionally
defined by
\begin{equation}
\!\not\!S_i{}^j = -\,{\frac 12}\,\kappa_{ik}{}^{jk} + {\frac  18}\,\kappa
\,\delta_i^j,\qquad \alpha = {\frac 1{12}}\,\kappa 
= {\frac 1{12}}\,\kappa_{ij}{}^{ij}.\label{Salpha}
\end{equation}
The {\it principal\/} (or the metric-dilaton) part
$^{(1)}\kappa_{ij}{}^{kl}$ of the constitutive tensor with 20
independent components will eventually be expressed in terms of the
metric (thereby cutting the 20 components in half).

It is straightforward to see that the axion $\alpha$ does not contribute
to the extended Fresnel equation at all (although they can produce effects
beyond the geometric optics approximation \cite{Carrol90,Yakov1}), whereas 
for a vanishing principal part $^{(1)}\kappa_{ij}{}^{kl} =0$, there is no 
wave propagation. In general, a nonvanishing skewon yields a very complicated 
picture for the wave surface \cite{skewonW}, i.e., a highly nontrivial 
birefringent wave propagation. 

One can prove \cite{lightcone} that by requiring the vanishing of 
birefringence, the quartic Fresnel wave surface reduces to a unique light 
cone. The latter defines, up to an overall conformal factor, a metric tensor 
$g^{ij}(x)$ with the correct Lorentzian signature, which is constructed 
directly in terms of the constitutive tensor $\kappa_{ij}{}^{kl}$.

\subsection{Dilaton and axion} 

Introducing the (Levi-Civita) dual of the excitation, $\check{H}^{ij}
:={\frac 12}\,\epsilon^{ijkl}\,H_{kl}$, we can finally rewrite the
spacetime relation for {\it vanishing birefringence} in vacuum as
\begin{equation}
  \check{H}^{ij}=[\underbrace{\lambda(x)}_{\hbox{dilaton}}
  \!\!\sqrt{-g}\,g^{ik}(x)\,g^{jl}(x)+\underbrace{\alpha(x)
    }_{\hbox{axion}} \epsilon^{ijkl}\: ]\,F_{kl}\,,\label{nobirefr}
\end{equation}
that is, we are left with the constitutive fields dilaton $\lambda$,
metric $g^{ij}$, and axion $\alpha$. The combination
$\sqrt{-g}\,g^{i[k}(x)\,g^{l]j}(x)$ is conformally invariant. 
In exterior calculus, (\ref{nobirefr}) reads
\begin{equation}
H=\lambda(x)\,^\star \!F+ \alpha(x)\,F\,,\label{H*F}
\end{equation}
with the Hodge star operator $^\star$. In turn, the Lagrangian of the 
electromagnetic theory, including dilaton and axion, reads
\begin{equation}\label{Lagr}
V = -\,{\frac 12}\left(\lambda\,F\wedge{}^\star\!F + \alpha\,F\wedge 
F\right)\,.
\end{equation}

\subsection{Light cone: null geodesics for a photon}

Inserting the spacetime relation
(\ref{nobirefr}) into (\ref{G4}), we find that the quartic Fresnel surface
(\ref{Fresnel}) reduces to the unique light cone, $g^{ij}q_iq_j = 0$. By 
taking a covariant derivative of the latter, we have
\begin{equation}
q^i\nabla_k q_i = q^i\nabla_i q_k = 0, \label{null}
\end{equation}
where we used the fact that $\nabla_{[i}q_{j]} = \partial_{[i}\partial_{j]}\Phi
\equiv 0$. Thus, we verified that the photon propagates along a null geodesic
(\ref{null}), in accordance with the equivalence principle. 

\section{Experimental tests}

\subsection{No birefringence} 

In vacuum, the propagation of light does not reveal any evidence of
birefringence \cite{chargenoncons,clausannalen}. Hence, one can
conclude that, up to a certain limit, there is no skewon field, and
the principal part of the constitutive tensor reduces to the standard
Maxwell-Lorentz form. The detailed analysis of the experimental limits
on the birefringence in vacuum can be found in
\cite{Kostelecky,Carrol90}.

\subsection{No axion?} 

As yet, the Abelian axion has not been found experimentally 
\cite{Sikivie,Wilczek1,Wilczek2,Lue99,Maj+Sen,Ni99}, see also the 
discussion of Cooper and Stedman \cite{Cooper} on corresponding ring 
laser experiments. However, the recent PVLAS experiments 
\cite{PVLAS,Biswas,Adler} pointed to some optical activity of the
vacuum provided an external magnetic field was present. A nontrivial axion 
field is considered as one of the possible explanations of these observations.
However, the interpretation of these experiments is still under discussion 
and the situation is not completely clear. Probably more independent 
experiments are needed. 

\subsection{No dilaton?} 

There is no direct evidence of the dilaton as yet, 
i.e., under normal circumstances, the dilaton seems to be a
constant field and thereby sets a certain scale, i.e., $\lambda(x)=
\lambda_{0}$, where $\lambda_{0}$ is the admittance of free space 
the value of which is, in SI-units, $\lambda_0\cong 1/(377\;\Omega)$.

However, the possibility of time and space variations of the fundamental
constants is discussed in the literature both from an experimental and
a theoretical point of view.
Of particular interest are certain indications that the fine
structure constant may slowly change on a cosmological time scale.
Ni \cite{Ni84} points out that the variation of fundamental constants
can be viewed as an indication of the violation of Einstein's equivalence
principle. 

Maxwell's equations follow from charge and flux conservation. Any
charge is proportional to the elementary charge $e$, any flux 
proportional to the elementary flux $h/e^2$. Consequently, if the 
electron charge $e$ and the Planck constant $h$ keep their values 
constant (independent of time, e.g.), then the quantities proportional 
to them, or any power of them, namely $e^{n_1}h^{n_2}$, with $n_1$ and 
$n_2$ as integer numbers, are also conserved. Therefore the time 
independence of $e$ and $h$ are the raisons d'\^{e}tre of the Maxwell 
equations. Or the other way round: If we want to uphold the Maxwell 
equations and thus QED, then we have to demand 
$e=\mbox{const}$ and $h=\mbox{const}$. 

However, there is a different possibility. In order to recognize this, let
us have a closer look at the definition of the fine structure constant:
\begin{equation}\label{fine}
\alpha_{\rm f} = {\frac {e^2}{2\varepsilon_0\,c\,h}} = 
{\frac {e^2}{2\,h\,\lambda_0}} = {\frac {\Omega_0}{2R_{\rm K}}}.
\end{equation}
As we can see, the fine structure constant is explicitly given in terms of
the ratio of two resistances --- vacuum impedance $\Omega_0 = 1/\lambda_0$ 
and von Klitzing constant $R_{\rm K}$ (the quantum Hall resistance). 
This relation is valid in {\it all} systems of units. Note that
the speed of light $c$ {\it disappeared completely!} Its ``presence"
in the first equality is in fact misleading (it is multiplied by 
$\varepsilon_0$!) and the proper
understanding is suggested only in the second equality where
$\lambda_0$ shows up instead, together with $e$ and $h$.


In other words, the formula (\ref{fine}) demonstrates that of the two
fundamental constants $\Omega_0$ and $c$ of electrodynamics, which appear 
naturally in the Maxwell-Lorentz electrodynamics (see the previous section), 
it is the {\it vacuum impedance} $\Omega_0$ which enters the fine structure 
constant and {\it not} the speed of light.

Recall now the argument \cite{Peres} that $e$ and $h$,
being 4D scalars, should not change in time and space provided one
wants to uphold the validity of the Maxwell equations (\ref{Maxwell}).
Then the variation of the fine structure constant $\alpha_{\rm f} =
\alpha_{\rm f}(t)$ forces us to conclude that $\lambda =
\lambda(t)$. An inspection of the Maxwell Lagrangian (\ref{Lagr})
then shows that $\lambda$ becomes a dynamical {\it dilaton\/} field.
In the axiomatic premetric approach to electrodynamics, we have a
scalar and a pseudoscalar part of the spacetime relation that are 
independent of the spacetime metric: these are the dilaton $\lambda$ 
and the axion $\alpha$. The variability of the fine structure constant 
thus may be explained without changing the Maxwell equations and QED 
by the presence of the {\it dilaton field} in a generalized  Maxwell-Lorentz 
spacetime relation.

\section{Conclusions}

We demonstrated here that the premetric axiomatic approach to 
electromagnetic theory provides a natural framework for the test 
of the equivalence principle. Which particular constitutive
(spacetime) relation $H=H(F)$ is valid? This is the central question. 
A complete analysis of this
question can be performed for the case of the local linear spacetime
relation, for which an extended Fresnel equation (\ref{Fresnel}) is 
derived in terms of the constitutive tensor $\kappa$. At present, there 
is no experimental evidence for the existence of birefringence for 
electromagnetic waves in vacuum. This fact alone reduces the structure
of the constitutive tensor to the simple form (\ref{nobirefr}) (or,
equivalently, to the spacetime relation (\ref{H*F})), introducing the 
(optical) spacetime metric $g$ and its scalar and pseudoscalar 
companions, dilaton and axion. A possible indication that the two 
latter premetric fields are nontrivial, in particular a dilaton $\lambda$,
can come from the possible variation of the fundamental physical constants. 
As to the axion, when $\alpha=$const, it does not affect the wave
propagation. However, as pointed out in \cite{aximeas,Ni05}, one may 
observe measurable effects at the interfaces between spacetime domains 
with different constant values of $\alpha$.

{\bf Acknowledgments}. This work was done with the support of the
DFG (Project He~528/21-1), Bonn.

\end{document}